\documentclass[10pt,twocolumn,superscriptaddress,
amssymb,amsmath,aps,prl]{revtex4-1}

\usepackage{graphicx}

\begin{document}

\setcitestyle{super}

\title{Radiation pressure measurement using a macroscopic oscillator in 
an ambient environment}
\date{November 24}
\author{Mikko Partanen}
\affiliation{Photonic Device Physics Laboratory, Department of Physics, 
Yonsei University, 50 Yonsei-ro Seodaemun-gu, Seoul 03722, Korea}
\affiliation{Photonics Group, Department of Electronics and Nanoengineering,
Aalto University, P.O. Box 15500, 00076 Aalto, Finland}
\author{Hyeonwoo Lee}
\affiliation{Photonic Device Physics Laboratory, Department of Physics, 
Yonsei University, 50 Yonsei-ro Seodaemun-gu, Seoul 03722, Korea}
\author{Kyunghwan Oh}
\affiliation{Photonic Device Physics Laboratory, Department of Physics, 
Yonsei University, 50 Yonsei-ro Seodaemun-gu, Seoul 03722, Korea}

\begin{abstract}
In contrast to current efforts to quantify the radiation pressure 
of light using 
nano-micromechanical resonators in cryogenic conditions, we proposed and 
experimentally demonstrated the radiation pressure measurement in ambient 
conditions by utilizing a macroscopic mechanical longitudinal oscillator with 
an effective mass of the order of $20$ g. The light pressure on a mirror 
attached to the oscillator was recorded in a Michelson interferometer and results 
showed, within the experimental accuracy of $3.9\%$, a good agreement with the 
harmonic oscillator model without free parameters.
\end{abstract}

\maketitle

\section{Introduction}
\vspace{-0.2cm}

According to Newton's second law, the force $\mathbf{F}$ on an object 
is 
well known to be equal to the rate of change of the momentum $\mathbf{p}$ of 
the object as
$\mathbf{F}=d\mathbf{p}/dt$.
From this fundamental law, we can expect the largest conversion of the 
optical momentum to the mechanical momentum of a medium when light is fully 
reflected from a mirror. The magnitude of this force on an ideal 100\% 
reflecting mirror in a vacuum is given by
\begin{equation}
 F=\frac{2P}{c},
 \label{eq:radiationpressureforce}
\end{equation}
where $P$ is the optical 
power and $c$ is the speed of light. If light is irradiated on a non-planar or partly absorbing object, the value of the radiation pressure is always smaller than its maximum value given in Eq.~\eqref{eq:radiationpressureforce}. Thus, the measurement of the maximum value of the optical force-power ratio, $F/P=2/c$, which has a universal value given in terms of the speed of light, provides a good test for the accuracy of the radiation pressure measurements.

The radiation pressure of light was first theoretically described by 
Maxwell \cite{Maxwell1873} in 
1873, and then experimentally measured 
independently
by Lebedev \cite{Lebedev1901} and by Nichols and Hull 
\cite{Nichols1903} in 1901, but the accuracy of these early experiments was 
very limited.
Despite being a century-old discovery, the
radiation pressure continues to be one of the key research interests in current 
optomechanics, such as in cooling of mechanical resonators 
\cite{Aspelmeyer2014,Chan2011,Gigan2006,Kleckner2006,Schliesser2006}, solar 
sail development \cite{Johnson2011}, ultra-high laser 
power measurements \cite{Williams2013,Pinot2019}, and nano-scale cantilevers' 
spring constant calibration \cite{Wilkinson2013,Weld2006}, to name a few.
Recently, there also has been renewed interest in the centennial 
Abraham-Minkowski controversy on the light momentum in a dielectric medium 
\cite{Leonhardt2006a,Partanen2017c,Bliokh2017a,Partanen2019a,
Partanen2019b,Bliokh2017b,Partanen2017e,Choi2017,Pozar2018,Partanen2018a,
Partanen2018b}.

The main trend in light pressure studies in recent years has been to 
miniaturize 
a mechanical oscillator to the nano-micro scale for a higher sensitivity to the 
radiation pressure \cite{Aspelmeyer2014,Gigan2006,Kleckner2006,Ryger2018}. 
However, 
optical forces in those nano-micromechanical systems have been directly 
accompanied by photothermal effects due to short thermal time constants of the 
miniaturized resonators 
\cite{Gigan2006,Kleckner2006,Metzger2004,Marti1992,Allegrini1992,Shaw2019}, 
which has 
required further sophisticated techniques to discern them from the radiation 
pressure effects. Therefore, various optical, mechanical, and thermal techniques 
have been developed to overcome the trade-off between the radiation pressure 
and 
the photothermal effects, such as complex resonator designs consisting of highly 
reflective multilayer coatings deposited on the cantilever to further increase 
the reflectivity \cite{Wilkinson2013,Kleckner2006}, attachment of an additional 
mass to increase the thermal time constant of the cantilever \cite{Weld2006}, 
or 
other ways to separate the optical force from the photothermal effects 
\cite{Ma2015,Ma2018}.

\begin{figure*}
\centering
\includegraphics[width=0.98\textwidth]{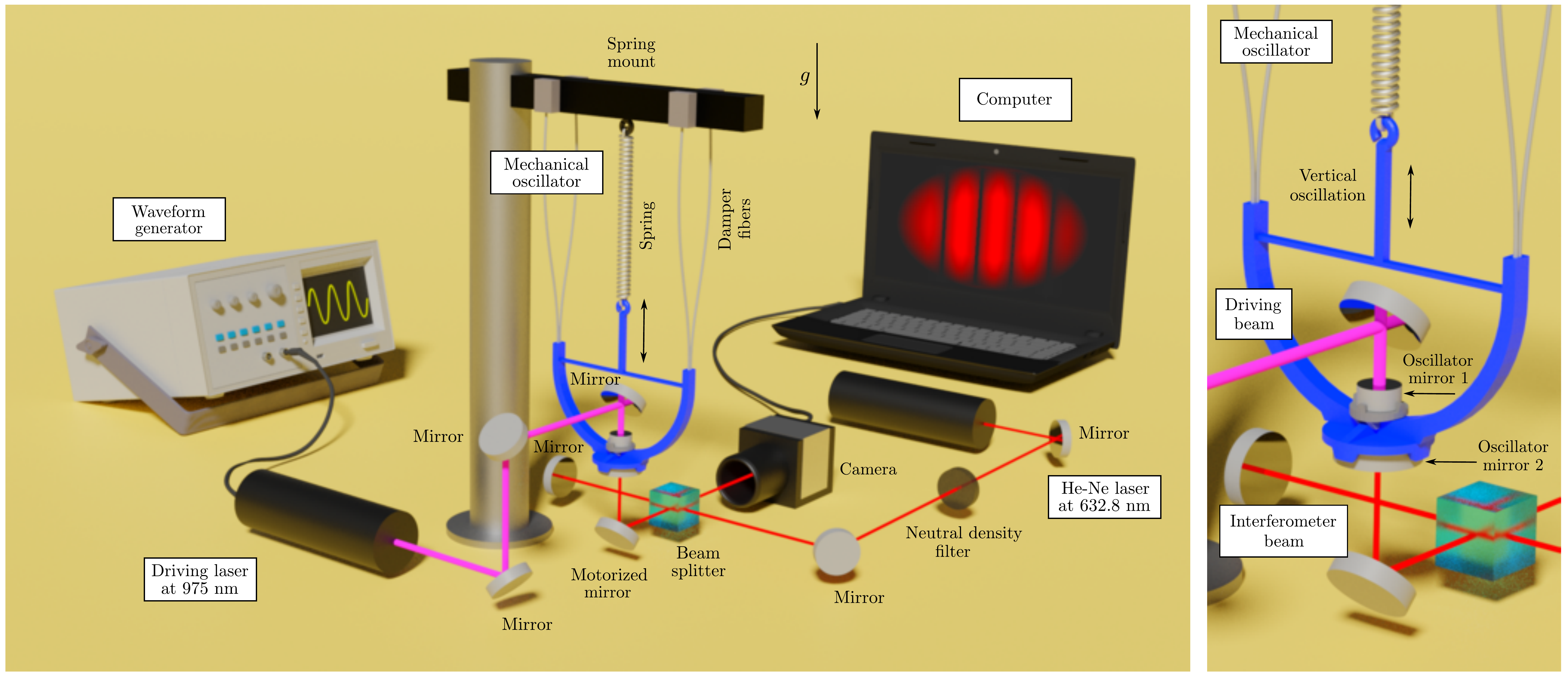}
\caption{\label{fig:setup}
The experimental setup consists of three 
main parts shown in the left panel: the mechanical 
oscillator, the diode laser driving the oscillator, and the Michelson 
interferometer.  The right 
panel focuses on the vertically hanging mechanical oscillator. 
The mechanical oscillator is driven by a modulated laser beam 
at 
975 nm through the reflection from the highly reflective 
oscillator mirror 1. The nanoscale oscillation is detected 
through the oscillator mirror 2 by 
the Michelson interferometer using the He-Ne laser at 632.8 nm.
The
motorized mirror below the oscillator is used for the remote 
tuning of the interference 
fringe spacing, but it is not actively controlled during the measurements. The 
illustration includes the damper fibers that are used for increasing the 
damping constant of the higher damping oscillator. The 
apparatuses are 
mounted on an actively damped optical table for isolating the setup from 
external acoustic and seismic vibrations. The oscillator part 
of the setup is also protected with plastic walls not shown in this 
illustration.
A more complete description of the 
experiment is presented in the Methods section. (Image created by using Blender 2.8, https://www.blender.org/, and Inkscape 0.92, https://inkscape.org/.)}
\vspace{-0.3cm}
\end{figure*}

In this work, we attempted a new direction opposite to the current trends 
by achieving a quantitative measurement of the radiation pressure of light in 
an ambient environment at room temperature by utilizing a macroscopic 
mechanical harmonic oscillator, which is orders of magnitude heavier than 
oscillators in previous reports 
\cite{Komori2020,Ma2015,Ma2018,Wagner2018,Evans2014,Melcher2014}. The 
experimental setup is illustrated in Fig.~\ref{fig:setup}. In 
contrast to conventional torsional oscillators used in most of the previous 
measurements, our oscillator is designed to allow only the longitudinal motion. 
Here we varied the mass and the damping constant to verify the accuracy of the 
harmonic oscillator model in the radiation pressure measurements. Note that our 
method can obviate the elaborated process to calibrate the spring constants 
\cite{Ma2015,Ma2018}, as the only additional measurement of the oscillator 
parameters is the 
direct determination of the oscillator masses using a digital scale.

\vspace{-0.2cm}
\section{Mechanical oscillator}
\vspace{-0.2cm}

The oscillator is driven optically by the reflection of the modulated laser 
beam 
at the wavelength of 975 nm at a  highly reflective dielectric 
mirror, 
which is the oscillator mirror 1 in Fig.~\ref{fig:setup}. The laser diode at 
this wavelength has
been widely used to pump optical amplifiers and provides a
good power stability of $\sim\pm0.5\%$.
The effective total
reflectivity of the oscillator mirror 1 was larger than 99.9\% (see Methods) and
Eq.~\eqref{eq:radiationpressureforce} can be used to quantify the optical 
force. The longitudinal displacement of the oscillator was detected by a Michelson interferometer using another laser at the wavelength of 632.8 nm. 
The interferometer beam was reflected from the oscillator 
mirror 2. The shifts of 
the interference fringes were recorded using a camera at a frame rate of 200 
frames per second, from which the displacement of the oscillator was estimated 
for various incident light powers. It is noteworthy, in particular, that 
photothermal effects can be excluded since light is reflected from a highly 
reflective dielectric mirror on a macroscopic mechanical oscillator whose 
thermal time constant is much longer than the modulation time of the laser 
field. A more complete description of the experimental setup is presented in 
the Methods section.

Newton's equation of motion for the mechanical oscillator with an effective 
mass 
$m$ is given by 
\cite{Aspelmeyer2014}
\begin{equation}
 \frac{d^2x}{dt^2}+2\zeta\omega_0\frac{dx}{dt}+\omega_0^2x=\frac{F}{m},
 \label{eq:oscillator}
\end{equation}
where $\omega_0$ is the undamped
resonance frequency of the harmonic oscillator, $\zeta$ is the damping 
coefficient, and $F$ is the net
external force. The damping coefficient is related to the Q factor as 
$Q=1/(2\zeta)$. When the mass of the vertically aligned spring 
is not negligible, the 
effective mass of the oscillator is given by 
$m=m_0+m_\mathrm{s}/3$, where $m_0$ is the rest mass of the oscillator and 
$m_\mathrm{s}$ is the rest mass of the spring \cite{Thomson1998}.

\begin{figure*}
\vspace{-0.1cm}
\centering
\includegraphics[width=0.83\textwidth]{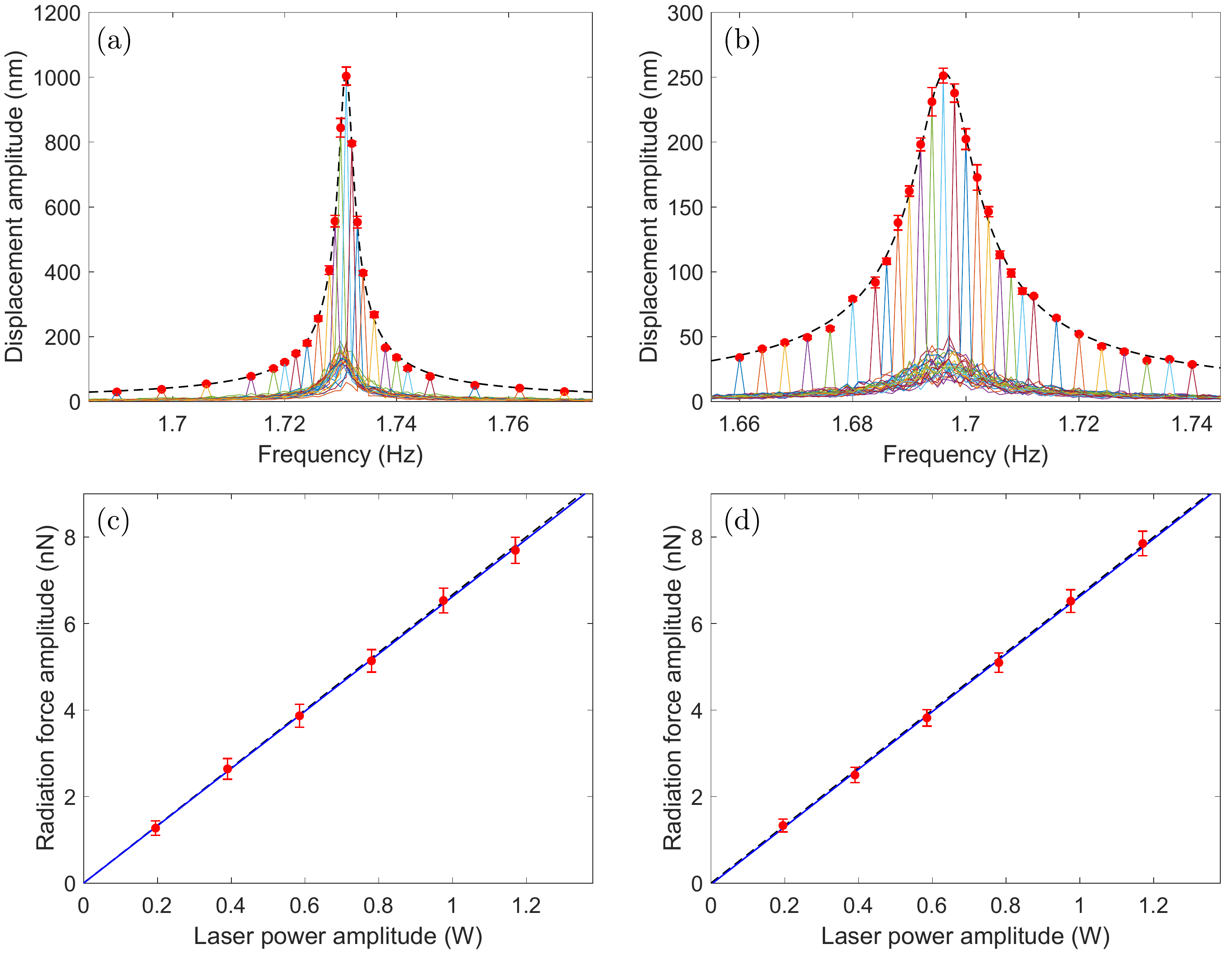}
\vspace{-0.2cm}
\caption{\label{fig:results}
The measured displacement amplitude of the mechanical oscillator is plotted 
(a) for the lower damping oscillator and (b) for the 
higher damping oscillator as a 
function of the modulation frequency of the driving laser with 
an example peak to 
peak power amplitude of 
$P_0=0.975$ W. Each graph that is marked with a solid line is the 
averaged 
frequency spectrum of a measurement for a single modulation frequency.
The graphs peak at the modulation frequency and the peak points are
marked with red dots. The modulation frequency is varied around the resonance 
frequency of the mechanical harmonic oscillator. The peak points of the graphs 
form together a curve that is the response function of the mechanical harmonic 
oscillator. The fitted harmonic oscillator response function is marked 
with the dashed line. In (c) and (d), the measured peak to peak radiation force 
amplitude is plotted for the two oscillators as a function of the peak to peak 
laser power amplitude. The least-squares regression lines are marked with the 
solid lines. The linear theoretical curve $F_0=2P_0/c$ is presented by the 
dashed 
lines. The horizontal error bars corresponding to the 
$\pm0.5\%$ uncertainty of the laser power are not shown because of their 
smallness.}
\vspace{-0.4cm}
\end{figure*}

If the force is harmonically modulated with the angular 
frequency $\omega$ as $F=F_0\cos^2(\frac{1}{2}\omega 
t)=\frac{1}{2}F_0[1+\cos(\omega t)]$, where $F_0$ is the peak to peak amplitude 
of the force, then the steady-state solution of 
Eq.~\eqref{eq:oscillator} is given as
$x(t)=x(\omega)\cos(\omega t+\varphi)+F_0/(2m\omega_0^2)$, where 
$\varphi=\arctan[2\omega\omega_0\zeta/(\omega^2-\omega_0^2)]\in[-\pi,0]$ and 
the displacement amplitude $x(\omega)$ is given by
\begin{equation} 
x(\omega)=\frac{F_0/m}{2\sqrt{
(2\omega\omega_0\zeta)^2+(\omega^2-\omega_0^2)^2}}.
\label{eq:displacement}
\end{equation}
The resonance frequency of the significantly underdamped oscillator with 
$\zeta<1/\sqrt{2}$ is $\omega_\mathrm{r}=\omega_0\sqrt{1-2\zeta^2}$ 
\cite{Thomson1998}. At this 
frequency, the displacement amplitude of the oscillator in 
Eq.~\eqref{eq:displacement} 
obtains its peak value, given by
$x_0=F_0/(4m\omega_0^2\zeta\sqrt{1-\zeta^2})$. Thus, from the measured peak 
value of the displacement amplitude, we can obtain the optical force 
as
\begin{equation}
 F_0=4m\omega_0^2\zeta\sqrt{1-\zeta^2}\,x_0.
 \label{eq:forceamplitude}
\end{equation}
Here, for our macroscopic 
oscillator, the undamped angular frequency and the damping constant can be 
accurately determined based on the position and width of the mechanical 
resonance peak and the effective mass of the oscillator can be 
determined from the oscillator and spring masses measured with a 
digital scale. The effective mass of the lower damping oscillator without the 
damper fibers in Fig.~\ref{fig:setup} is $m=(18.363\pm 0.001)$ g, while the 
effective mass the higher damping oscillator with the damper fibers is 
$m=(19.007\pm 0.001)$ g. Note that the difference in the oscillator masses is 
mainly produced in their fabrication and it is not due to the damper fibers 
whose total mass is less than 0.2 g. The damper fiber is
commercially available optical fiber (Thorlabs, FG105LCA), 
which provides a very high mechanical stability against tensile stress.

\vspace{-0.2cm}
\section{Results}
\vspace{-0.2cm}

Figure \ref{fig:results} presents the experimental results. In 
Fig.~\ref{fig:results}(a), the measured displacement amplitude of the lower 
damping oscillator is plotted as a function of the modulation frequency of 
the driving laser field with an example peak to peak 
power amplitude of 
$P_0=0.975$ W. Fig.~\ref{fig:results}(b) presents the same plot for 
the higher 
damping oscillator. Each graph that is marked with a solid line is the 
averaged frequency spectrum of a measurement for a single modulation 
frequency. The measurement time is an integer multiple of the modulation period 
close to 1000 s and the ensemble averaging is made over 10 or more 
measurements. The error in the displacement amplitude is the standard deviation 
of the average and it could be made smaller by averaging over a larger number 
of measurements.

In Figs.~\ref{fig:results}(a) and \ref{fig:results}(b), one can see that the 
fitted harmonic oscillator response function in Eq.~\eqref{eq:displacement} 
accurately describes the experimental results of both oscillators. It can be 
noted that, in the presense of photothermal effects, this response function 
would be modified from the ideal harmonic oscillator form as described, e.g., 
in Refs.~\cite{Ma2015,Ma2018}. Thus, the ideal harmonic oscillator form of the 
frequency response function in Fig.~\ref{fig:results}(a) indicates that 
photothermal effects are negligible in our macroscopic setting as expected. In 
Figs.~\ref{fig:results}(a) and \ref{fig:results}(b), one can also see that the 
mechanical resonance peak is observable in the noise spectrum that is seen 
below 
the fitted harmonic oscillator response function.

\begin{table*}
 \setlength{\tabcolsep}{5.0pt}
 \renewcommand{\arraystretch}{1.4}
 \caption{\label{tbl:comparison}
 Comparison of the orders of magnitudes of characteristic 
physical quantities, i.e., mechanical frequency $f_0$, effective oscillator 
mass 
$m$, laser power modulation amplitude $P_0$, peak displacement amplitude $x_0$, 
force amplitude
$F_0$, and quality factor $Q$, in the present and selected previous works on 
the measurement of optical forces with mechanical oscillators. For direct 
comparability, the quantities in the table are obtained by fitting the harmonic 
oscillator response function of Eq.~\eqref{eq:displacement} to the experimental 
data corresponding to Fig.~\ref{fig:results}(a) or \ref{fig:results}(b) of 
the present work. In the column for the optical force per power, $F_0/P_0$, 
the uncertainties do not account for the uncertainties in the determination of 
the optical power, which is separately shown in the last column.}
\vspace{0.2cm}
\begin{tabular}{lllllllll}
   \hline\hline
Work & $f_0$ (Hz) & $m$ (kg) & $P_0$ (W) & $x_0$ (m) & $F_0$ (N) & $Q$ & 
$F_0/P_0$ (1/$c$) & $\Delta P_0/P_0$ \\[2pt]
   \hline
Present & $1$ & $10^{-2}$ & $1$ & $10^{-7}$ & $10^{-8}$ & $10^2$ & 
$1.998\pm0.077$ $^\mathrm{a}$ & $0.5\%$ \\
Weld \emph{et al.}~2006 \cite{Weld2006} & $10^2$ & $10^{-9}$ & $10^{-6}$ & 
$10^{-9}$ & $10^{-14}$ & $10^4$ & & \\
Wilkinson \emph{et al.}~2013 \cite{Wilkinson2013} & $10^2$ & $10^{-7}$ & 
$10^{-3}$ & 
$10^{-9}$ & $10^{-11}$ & $10^3$ & $1.974\pm0.116$ $^\mathrm{b}$ & $1\%$ \\
Wagner \emph{et al.}~2018 \cite{Wagner2018} & $10^3$ & $10^{-8}$ & $10^{-4}$ & 
$10^{-11}$ & $10^{-10}$ & $10^3$ & $1.656\pm0.043$ $^\mathrm{c}$ & $8\%$ \\
Melcher \emph{et al.}~2014 \cite{Melcher2014} & $10^4$ & $10^{-9}$ & $10^{-3}$ 
& 
$10^{-10}$ & $10^{-13}$ & $10^3$ & & \\
Kleckner \emph{et al.}~2006 \cite{Kleckner2006} & $10^4$ & $10^{-11}$ & 
$10^{-3}$ & 
$10^{-7}$ & $10^{-12}$ & $10^5$ & $2.113\pm0.163$ $^\mathrm{d}$ & $20\%$ \\
Ma \emph{et al.}~2018 \cite{Ma2018} & $10^4$ & $10^{-12}$ & $10^{-4}$ & 
$10^{-10}$ & $10^{-12}$ & $10$ & $1.234\pm0.235$ $^\mathrm{e}$ & $3\%$ \\
Ma \emph{et al.}~2015 \cite{Ma2015} & $10^4$ & $10^{-13}$ & $10^{-3}$ & 
$10^{-8}$ & $10^{-11}$ & $10$ & $0.521\pm0.010$ $^\mathrm{f}$ & $5\%$ \\
Evans \emph{et al.}~2014 \cite{Evans2014} & $10^4$ & $10^{-14}$ & $10^{-5}$ & 
$10^{-10}$ & $10^{-13}$ & $10^2$ & & \\
Gigan \emph{et al.}~2006 \cite{Gigan2006} & $10^5$ & $10^{-11}$ & $10^{-3}$ & 
$10^{-10}$ & $10^{-11}$ & $10^4$ & & \\
\hline\hline
 \end{tabular}
 \scriptsize
 \setlength{\tabcolsep}{1.0pt}
 \begin{tabular}{lp{15cm}}
  $^\mathrm{a}$ & Determined from the slope of the line in 
Fig.~\ref{fig:results}(d).\\
  $^\mathrm{b}$ & Determined as $F_0/P_0$, where $F_0=kA_0$ with $k$ being the 
spring constant obtained from the nanoindenter and the off-resonance amplitude 
$A_0=2P_0/(m\omega_0^2c)$ is obtained from the quantities given in Table I of 
the reference.\\
 $^\mathrm{c}$ & Determined from the values in Table 1 of the reference as 
$F_\mathrm{sho}/P_\mathrm{c}$.\\
 $^\mathrm{d}$ & Determined as $F_0/P_0$, where $F_0=kA_\mathrm{rad}$ with 
given values for the spring constant $k$ determined from the thermal vibration 
spectrum and for the off-resonance amplitude $A_\mathrm{rad}$.\\
 $^\mathrm{e}$ & Determined from Fig.~3(b) of the reference at 750 nm. The 
value is much lower than $2/c$ due to the transmission and absorption 
which reduce the reflectivity.\\
 $^\mathrm{f}$ & Determined from the slope of the line fitted in the 
experimental data of Fig.~4 of the reference. The value is much lower than $2/c$ 
due to the partial transmission which reduces the reflectivity.
 \end{tabular}
 \vspace{-0.4cm}
\end{table*}

In the least-squares fitting of the harmonic oscillator response 
function in the experimental data of the lower damping oscillator in 
Fig.~\ref{fig:results}(a), the undamped 
frequency of the mechanical oscillator is found to be 
$f_0=(1.730943\pm 0.000018)$ Hz and the damping constant is found to be 
$\zeta=0.000750\pm 0.000012$, which 
corresponds to the quality factor of $Q=667\pm 11$. The errors indicate the 
68.27\% confidence intervals, which correspond to one standard deviation 
of normally distributed quantities.
Using the experimental data of the higher damping oscillator in 
Fig.~\ref{fig:results}(b), we respectively obtain the 
undamped oscillator
frequency of $f_0=(1.696275\pm 0.000073)$ Hz and the damping constant of 
$\zeta=0.003006\pm 0.000051$, which corresponds to $Q=166.3\pm 2.8$.

Figure \ref{fig:results}(c) shows the measured peak 
to peak radiation force amplitude of the lower damping oscillator 
following from Eq.~\eqref{eq:forceamplitude} as 
a function of the peak to peak laser power amplitude. The corresponding 
radiation force amplitude graph for the higher damping oscillator
is presented in Fig.~\ref{fig:results}(d). In both cases, the 
uncertainty in the laser 
power is $\pm0.5\%$. For 
the lower damping oscillator, the slope of the 
least-squares regression line is 
$dF_0/dP_0=(6.63\pm0.29)\times10^{-9}\;\text{s/m}=(1.986\pm0.086)/c$. 
The relative error is 4.3\%, from which 1.6\% comes from the determination of 
the damping constant in the fitting of 
Fig.~\ref{fig:results}(a) and 2.7\% comes from the determination of the peak 
displacement amplitude and the uncertainty of the laser power. 
These values include
uncertainties 
related to the laser power fluctuations around the expectation value. For the 
higher damping 
oscillator, the regression line 
is
$dF_0/dP_0=(6.66\pm0.26)\times10^{-9}\;\text{s/m}=(1.998\pm0.077)/c$, where 
the relative error is 3.9\%, 
from which 1.7\% comes from the determination of 
the damping constant and 2.2\% comes from the determination of the peak 
displacement amplitude and the uncertainty of the laser power.
The slope of the 
corresponding universal
theoretical line, $F_0=2P_0/c$, is $2/c=6.67\times10^{-9}$ s/m.
Thus, the experimental results 
of both the lower and higher damping oscillators
agree with the theory within the experimental accuracy.

That our results are in accurate agreement with the theory is another indication for the insignificance of thermal effects in the present experimental setup. However, the role of thermal effects could be studied in more detail in further experiments. For example, we could follow the closely related experiments by Po{\v z}ar \emph{et al.} \cite{Pozar2015} and repeat the experiment by using mirrors with increasingly higher reflectivities, starting from low reflectivities, in which case thermal effects are surely present, and continuing to ultra high reflectivities, in which case thermal effects become insignificant. We could also accurately verify that the results are independent of the excitation beam radius, as must be the case for true radiation pressure.

Comparison of the orders of 
magnitudes of the characteristic physical quantities in the present and 
selected 
previous works is presented in Table \ref{tbl:comparison}. Most notably, it is 
seen that the effective oscillator mass in the present 
work is several orders of magnitude larger than the oscillator masses in 
previous works. Also, the mechanical frequency is lower and the laser power 
modulation amplitude is larger compared to previous works. The comparison
of the optical force per power, $F_0/P_0$, shows that only a few works
have used the mechanical oscillator to study the force on an ideal 
mirror with $F_0/P_0=2/c$ and found an accurate correspondence with
this relation as we have done in the present work. Regarding 
the determination of the absolute radiation force, in contrast to our work, the 
main experimental uncertainties in most previous works have originated from the 
determination of the magnitudes of the small optical power and the spring 
constant of the oscillator.

\vspace{-0.2cm}
\section{Conclusions}
\vspace{-0.2cm}

In conclusion, we have have demonstrated that the radiation pressure of light 
can be accurately measured in ambient environment by utilizing a macroscopic 
mechanical 
oscillator and detecting how the modulation of the optical signal can be tuned 
to drive the nanoscale motion of the oscillator. We have carried out 
measurements for two oscillators with different masses and damping constants, 
and shown that the correspondence between the theory and experiment is obtained 
within the relative experimental accuracy. The introduced setup can also be 
used for probing 
optical forces when the oscillator is driven through the optical fibers that 
are present as the damper fibers in the setup of Fig.~\ref{fig:setup}, but 
these investigations related to the Abraham-Minkowski 
controversy are left as a topic of further work. One might also be able to access additional information on the momentum of light in a medium by immersing the oscillator or its driving mirror in liquids with known refractive indices. However, in the related analysis, one should account for the effects of fluid dynamics and also the surface tension if the immersion is only partial. Our macroscopic oscillator 
setup and its larger-scale variations can also be used for measuring high laser 
powers through the determination of the radiation pressure.

\vspace{-0.2cm}
\section{Methods}
\vspace{-0.2cm}

\noindent\textbf{Mechanical oscillator.}
The mechanical oscillator masses are fabricated by 3D printing from polylactic 
acid, commonly known as PLA. The printing is performed by using the fused 
deposition modeling (FDM) technique. The design of the oscillator mass 
includes two mirror mounts
and the 
hook that connects the oscillator mass to the mechanical 
extension spring, which carries the weight of the oscillator mass.
The semicircle form of the oscillator has a mean radius of $R=4.25$ cm.
The design of the higher damping oscillator mass also
includes the damper fibers (Thorlabs, 
FG105LCA) whose ends are clued to the oscillator.
The rest masses of the lower and higher damping mechanical oscillators are 
15.606 g and 16.250 g, respectively. These masses include 1.637 g of the mass 
of 
the oscillator mirror 1 (Edmund Optics, 63-129) and
6.715 
g of the mass 
of the oscillator mirror 2 (Thorlabs, BB1-E02).

\vspace{0.2cm}\noindent\textbf{Driving laser.}
The driving laser beam at 975 nm is generated by a multimode laser diode module 
(Lumics, LU0975T090) and it is directed through a low-loss 0.22 NA silica core 
multimode fiber (Thorlabs, 
FG105LCA) to the collimation arrangement after which the laser beam hits the 
mirror on the 
mechanical oscillator.
The intensity of the driving 
laser beam is modulated by 
the waveform generator (Agilent, 33120A) connected to the laser driver 
(Arroyo Instruments, LaserPak 485-08-05). The temperature of the laser is 
controlled with the temperature controller (Arroyo Instruments, TECPak 
585-04-08). The laser 
beam coming out from the end of the multimode fiber is collimated by using the
achromatic lens (Thorlabs, AC050-008-B) with a diameter of 5 mm and focal 
length of 7.5mm. After the collimating lens, we use 
two mirrors (Thorlabs, BB1-E03) for accurately aligning the beam normally to 
the oscillator mirror 1. In addition, at 80 cm 
before 
the mechanical oscillator, we use a focusing lens to adjust the spot 
size of the laser beam so that the spot diameter is 5 mm when it 
hits the mirror on the mechanical oscillator. The focusing and diverging 
angles of the order 
of one percent are so small that the non-normally 
incident field components on the mechanical oscillator have negligible 
influence on our measurement results. Thus, we can safely use the normal 
incidence assumption in the analysis of the experimental results.
We also minimize thermal drifts 
by 
using a highly reflective dielectric oscillator mirror 1
with an effective total reflectivity larger than $99.9$\%. This effective total reflectivity consists of the specular reflectivity $>99.8$\% and the estimated normal component of the diffusive reflectivity corresponding to more than half of the light that is not specularly reflected. The reflectivity of this mirror is 
assumed unity in the analysis, in which case Eq.~\eqref{eq:radiationpressureforce} can be used for the 
optical force.

\vspace{0.2cm}\noindent\textbf{Laser power measurement.}
The laser power 
is measured using an optical power meter (Thorlabs, PM400) with an integrating 
sphere sensor (Thorlabs, S145C) at the position 
before the laser beam hits the last mirror directing the beam toward the 
mechanical oscillator. The reduction of the power due to the reflectivity 
of 
99.5\% of the last mirror is accounted for in the analysis.
The peak to peak amplitude of the sinusoidal modulation was 2.0\% smaller than the otherwise stationary beam.
Thus, for example, for a stationary beam with 
a power of 1.000 W, after adding the modulation and accounting for the 
reflectivity of the last mirror, the peak to peak power hitting the 
oscillator mirror 1 becomes 0.975 W, which is the value used in the measurements corresponding to Figs.~\ref{fig:results}(a) and \ref{fig:results}(b).

\vspace{0.2cm}\noindent\textbf{Extension springs.}
In the experiments, we have used three hard extension springs in series to 
obtain a suitably small total spring constant. The upper and lower springs 
(Acxess Spring, 
PE016-312-129000-MW-2500-MH-N-IN) both have a mass of 
$m_\mathrm{s,1}=3.025$ 
g and
a reported extension rate of $k_\mathrm{s,1}=5$ N/m, while the middle spring 
(Acxess Spring, 
PE016-312-90250-MW-1880-MH-N-IN) has a mass of $m_\mathrm{s,2}=2.221$ g and a 
reported extension rate of 
$k_\mathrm{s,2}=7$ N/m. The springs 
are made of music wire and 
they have cross-over type hooks at their ends.
The total mass of the three springs in series 
is given by $m_\mathrm{s}=8.271$ g. In 
addition to the extension rates above, the masses of 
the vertically aligned springs also contribute to the 
total spring constant of the oscillator. In the case of the higher damping 
oscillator, where damper fibers are used, there is 3.0 mm loose in the damper 
fibers 
so 
that the oscillator mass is entirely carried by the spring. In our analysis of 
the experimental results, we 
use the effective mass, undamped angular frequency and the damping constant as 
the only parameters of the oscillator. Thus, the total spring constant of the 
system can be determined from the experimental results as 
$k=m\omega_0^2$. For the lower damping oscillator, we thus have $k=2.172$ N/m,
while for the higher damping oscillator we have $k=2.159$ N/m.

\vspace{0.2cm}\noindent\textbf{Michelson interferometer.}
The motion of the mechanical oscillator is detected 
by the Michelson interferometer to which the oscillator is connected by 
setting the oscillator mirror 2 in one 
of 
the 
two interferometer arms. One of the other interferometer arm mirrors is 
motorized so that it can be used for the remote tuning of the interference 
fringe spacing. The fringe spacing is adjusted only before the measurements and 
it is not actively changed during the measurements. The arm length of the 
interferometer is about 10 
cm.
All mirrors in the interferometer arms have a reflectivity of over 
99\% (Thorlabs, BB1-E02). The interferometer utilizes the 5-mW continuous-wave 
TEM$_{00}$ He-Ne 
laser (JDSU, 1125P) at 632.8 nm. The laser power is reduced by a factor of 1/10 
by a neutral density
filter (Thorlabs, NE10A). The dynamics of the interference fringes is recorded 
by a CMOS camera (Edmund Optics, EO-0413C) with a frame rate of 200 
frames per 
second. The frame size recorded is 600x30 pixels.

\vspace{0.2cm}\noindent\textbf{Tracking the dynamics of interference fringes.}
The dynamics of the interference fringes is tracked from the recorded video 
files by observing the movement of the intensity maxima and minima frame by 
frame. The interference fringes are illustrated in the computer screen of 
Fig.~\ref{fig:setup}, where the movement of the fringes takes place in 
the horizontal direction when other disturbances are settled down. If the 
fringes move a distance that is equal to the distance between two intensity 
maxima, this indicates that the mechanical oscillator moves half a wavelength 
in 
the vertical direction.
For efficient analysis of millions of frames, a C++ code was written that 
utilizes the Open Computer 
Vision Library (Open CV). The code also detects possible changes in the 
distance between intensity maxima as these scale changes indicate motion of the 
oscillator in lateral directions.
Since the mechanical oscillator is hanging on a spring, 
it can move in all three dimensions. However, the interferometer is the most 
sensitive 
for the vertical motion of the oscillator that is of our interest. If the 
oscillator is 
particularly disturbed, the interference fringes can also rotate and the scale 
of the fringes can 
vary. These effects are, however, negligibly small 
during measurement conditions, when 
external noise sources are minimized. The frequency responses of these effects 
are also
different from the mechanical resonance frequency of the oscillator so they 
do not contribute to the magnitude of the observed displacement amplitude.

\vspace{0.2cm}\noindent\textbf{Acoustic and seismic isolation.}
The apparatuses are mounted on an actively damped optical table for isolating 
the setup from acoustic and seismic vibrations. The mechanical oscillator 
part of the setup is covered with plastic walls to minimize air flows. The 
measurements were carried out at nighttime to minimize disturbances in the 
surroundings of the laboratory.

\noindent\textbf{Acknowledgements}\\
This work has been funded by European Union's Horizon 2020 Marie 
Sk\l{}odowska-Curie Actions (MSCA) individual fellowship 
DynaLight under Contract No.~846218 and the National Research Foundation of 
Korea (NRF) grant by the Korea government (MSIT) under Contract 
No.~2019R1A2C2011293.

\noindent\textbf{Author contributions statement}\\
M.P. and K.O. initiated the project; M.P. designed the experimental setup; H.L. and K.O. commented on the setup; M.P. and H.L. built the setup; M.P. carried out the measurements, analyzed the results, and wrote the main manuscript text; all authors participated in discussions over the results and revised the manuscript.

\noindent\textbf{Competing interests}\\
The authors declare no competing interests.


\vspace{-0.3cm}
\begin{thebibliography}{10}
\newcommand{\enquote}[1]{``#1''}

\bibitem{Maxwell1873}
J.~C. Maxwell, \emph{A Treatise on Electricity and Magnetism}, Oxford
  University, Oxford (1873).

\bibitem{Lebedev1901}
P.~N. Lebedev, \enquote{Experimental examination of light pressure,} \emph{Ann.
  Phys.} \textbf{6}, 433 (1901).

\bibitem{Nichols1903}
E.~F. Nichols and G.~F. Hull, \enquote{The pressure due to radiation,}
  \emph{Phys. Rev. (Series I)} \textbf{17}, 26 (1903).

\bibitem{Aspelmeyer2014}
M.~Aspelmeyer, T.~J. Kippenberg, and F.~Marquardt, \enquote{Cavity
  optomechanics,} \emph{Rev. Mod. Phys.} \textbf{86}, 1391 (2014).

\bibitem{Chan2011}
J.~Chan, T.~P.~M. Alegre, A.~H. Safavi-Naeini, J.~T. Hill, A.~Krause,
  S.~Gr\"{o}blacher, M.~Aspelmeyer, and O.~Painter, \enquote{Laser cooling of a
  nanomechanical oscillator into its quantum ground state,} \emph{Nature}
  \textbf{478}, 89 (2011).

\bibitem{Gigan2006}
S.~Gigan, H.~R. B\"{o}hm, M.~Paternostro, F.~Blaser, G.~Langer, J.~B.
  Hertzberg, K.~C. Schwab, D.~B\"{a}uerle, M.~Aspelmeyer, and A.~Zeilinger,
  \enquote{Self-cooling of a micromirror by radiation pressure,} \emph{Nature}
  \textbf{444}, 67 (2006).

\bibitem{Kleckner2006}
D.~Kleckner and D.~Bouwmeester, \enquote{Sub-kelvin optical cooling of a
  micromechanical resonator,} \emph{Nature} \textbf{444}, 75 (2006).

\bibitem{Schliesser2006}
A.~Schliesser, P.~Del'Haye, N.~Nooshi, K.~J. Vahala, and T.~J. Kippenberg,
  \enquote{Radiation pressure cooling of a micromechanical oscillator using
  dynamical backaction,} \emph{Phys. Rev. Lett.} \textbf{97}, 243905 (2006).

\bibitem{Johnson2011}
L.~Johnson, M.~Whorton, A.~Heaton, R.~Pinson, G.~Laue, and C.~Adams,
  \enquote{Nanosail-{D}: A solar sail demonstration mission,} \emph{Acta
  Astronaut.} \textbf{68}, 571  (2011).

\bibitem{Williams2013}
P.~A. Williams, J.~A. Hadler, R.~Lee, F.~C. Maring, and J.~H. Lehman,
  \enquote{Use of radiation pressure for measurement of high-power laser
  emission,} \emph{Opt. Lett.} \textbf{38}, 4248 (2013).

\bibitem{Pinot2019}
P.~Pinot and Z.~Silvestri, \enquote{Optical power meter using radiation
  pressure measurement,} \emph{Measurement} \textbf{131}, 109  (2019).

\bibitem{Wilkinson2013}
P.~R. Wilkinson, G.~A. Shaw, and J.~R. Pratt, \enquote{Determination of a
  cantilever's mechanical impedance using photon momentum,} \emph{Appl. Phys.
  Lett.} \textbf{102}, 184103 (2013).

\bibitem{Weld2006}
D.~M. Weld and A.~Kapitulnik, \enquote{Feedback control and characterization of
  a microcantilever using optical radiation pressure,} \emph{Appl. Phys. Lett.}
  \textbf{89}, 164102 (2006).

\bibitem{Leonhardt2006a}
U.~Leonhardt, \enquote{Momentum in an uncertain light,} \emph{Nature}
  \textbf{444}, 823 (2006).

\bibitem{Partanen2017c}
M.~Partanen, T.~H\"ayrynen, J.~Oksanen, and J.~Tulkki, \enquote{Photon mass
  drag and the momentum of light in a medium,} \emph{Phys. Rev. A} \textbf{95},
  063850 (2017).

\bibitem{Bliokh2017a}
K.~Y. Bliokh, A.~Y. Bekshaev, and F.~Nori, \enquote{Optical momentum, spin, and
  angular momentum in dispersive media,} \emph{Phys. Rev. Lett.} \textbf{119},
  073901 (2017).

\bibitem{Partanen2019a}
M.~Partanen and J.~Tulkki, \enquote{Lorentz covariance of the mass-polariton
  theory of light,} \emph{Phys. Rev. A} \textbf{99}, 033852 (2019).

\bibitem{Partanen2019b}
M.~Partanen and J.~Tulkki, \enquote{Lagrangian dynamics of the coupled
  field-medium state of light,} \emph{New J. Phys.} \textbf{21}, 073062 (2019).

\bibitem{Bliokh2017b}
K.~Y. Bliokh, A.~Y. Bekshaev, and F.~Nori, \enquote{Optical momentum and
  angular momentum in complex media: from the {A}braham-{M}inkowski debate to
  unusual properties of surface plasmon-polaritons,} \emph{New J. Phys.}
  \textbf{19}, 123014 (2017).

\bibitem{Partanen2017e}
M.~Partanen and J.~Tulkki, \enquote{Mass-polariton theory of light in
  dispersive media,} \emph{Phys. Rev. A} \textbf{96}, 063834 (2017).

\bibitem{Choi2017}
H.~Choi, M.~Park, D.~S. Elliott, and K.~Oh, \enquote{Optomechanical measurement
  of the {A}braham force in an adiabatic liquid-core optical-fiber waveguide,}
  \emph{Phys. Rev. A} \textbf{95}, 053817 (2017).

\bibitem{Pozar2018}
T.~Po\ifmmode\check{z}\else\v{z}\fi{}ar,
  J.~Lalo\ifmmode\check{s}\else\v{s}\fi{}, A.~Babnik,
  R.~Petkov\ifmmode\check{s}\else\v{s}\fi{}ek, M.~Bethune-Waddell, K.~J. Chau,
  G.~V.~B. Lukasievicz, and N.~G.~C. Astrath, \enquote{Isolated detection of
  elastic waves driven by the momentum of light,} \emph{Nat. Commun.}
  \textbf{9}, 3340 (2018).

\bibitem{Partanen2018a}
M.~Partanen and J.~Tulkki, \enquote{Mass-polariton theory of sharing the total
  angular momentum of light between the field and matter,} \emph{Phys. Rev. A}
  \textbf{98}, 033813 (2018).

\bibitem{Partanen2018b}
M.~Partanen and J.~Tulkki, \enquote{Light-driven mass density wave dynamics in
  optical fibers,} \emph{Opt. Express} \textbf{26}, 22046 (2018).

\bibitem{Ryger2018}
I.~{Ryger}, A.~B. {Artusio-Glimpse}, P.~{Williams}, N.~{Tomlin}, M.~{Stephens},
  K.~{Rogers}, M.~{Spidell}, and J.~{Lehman}, \enquote{Micromachined force
  scale for optical power measurement by radiation pressure sensing,}
  \emph{IEEE Sens. J.} \textbf{18}, 7941 (2018).

\bibitem{Metzger2004}
C.~H. Metzger and K.~Karrai, \enquote{Cavity cooling of a microlever,}
  \emph{Nature} \textbf{432}, 1002 (2004).

\bibitem{Marti1992}
O.~Marti, A.~Ruf, M.~Hipp, H.~Bielefeldt, J.~Colchero, and J.~Mlynek,
  \enquote{Mechanical and thermal effects of laser irradiation on force
  microscope cantilevers,} \emph{Ultramicroscopy} \textbf{42-44}, 345  (1992).

\bibitem{Allegrini1992}
M.~Allegrini, C.~Ascoli, P.~Baschieri, F.~Dinelli, C.~Frediani, A.~Lio, and
  T.~Mariani, \enquote{Laser thermal effects on atomic force microscope
  cantilevers,} \emph{Ultramicroscopy} \textbf{42-44}, 371  (1992).

\bibitem{Shaw2019}
G.~A. Shaw, J.~Stirling, J.~Kramar, P.~Williams, M.~Spidell, and R.~Mirin,
  \enquote{Comparison of electrostatic and photon pressure force references at
  the nanonewton level,} \emph{Metrologia} \textbf{56}, 025002 (2019).

\bibitem{Ma2015}
D.~Ma, J.~L. Garrett, and J.~N. Munday, \enquote{Quantitative measurement of
  radiation pressure on a microcantilever in ambient environment,} \emph{Appl.
  Phys. Lett.} \textbf{106}, 091107 (2015).

\bibitem{Ma2018}
D.~Ma and J.~N. Munday, \enquote{Measurement of wavelengthdependent radiation
  pressure from photon reflection and absorption due to thin film
  interference,} \emph{Sci. Rep.} \textbf{8}, 15930 (2018).

\bibitem{Komori2020}
K.~Komori, Y.~Enomoto, C.~P. Ooi, Y.~Miyazaki, N.~Matsumoto, V.~Sudhir,
  Y.~Michimura, and M.~Ando, \enquote{Attonewton-meter torque sensing with a
  macroscopic optomechanical torsion pendulum,} \emph{Phys. Rev. A}
  \textbf{101}, 011802 (2020).

\bibitem{Wagner2018}
R.~Wagner, F.~Guzman, A.~Chijioke, G.~K. Gulati, M.~Keller, and G.~Shaw,
  \enquote{Direct measurement of radiation pressure and circulating power
  inside a passive optical cavity,} \emph{Opt. Express} \textbf{26}, 23492
  (2018).

\bibitem{Evans2014}
D.~R. Evans, P.~Tayati, H.~An, P.~K. Lam, V.~S.~J. Craig, and T.~J. Senden,
  \enquote{Laser actuation of cantilevers for picometre amplitude dynamic force
  microscopy,} \emph{Sci. Rep.} \textbf{4}, 5567 (2014).

\bibitem{Melcher2014}
J.~Melcher, J.~Stirling, F.~G. Cervantes, J.~R. Pratt, and G.~A. Shaw,
  \enquote{A self-calibrating optomechanical force sensor with femtonewton
  resolution,} \emph{Appl. Phys. Lett.} \textbf{105}, 233109 (2014).

\bibitem{Thomson1998}
W.~T. Thomson and M.~Dahleh, \emph{Theory of Vibration with Applications},
  Prentice-Hall, New Jersey (1998).

\bibitem{Pozar2015}
T.~Po\v{z}ar, A.~Babnik, and J.~Mo\v{z}ina, \enquote{From laser ultrasonics to
  optical manipulation,} \emph{Opt. Express} \textbf{23}, 7978 (2015).

\end{thebibliography}
\end{document}